\documentclass{PoS}

\title{Permutation group $S_N$ and hadron spectroscopy}

\ShortTitle{Permutation group $S_N$ and hadron spectroscopy}

\author{\speaker{Dan Pirjol}\\    %\thanks{A footnote may follow.}\\
        Department of Particle Physics, National Institute for Physics and Nuclear Engineering,
        077125 Bucharest, Romania\\
        E-mail: \email{pirjol@mac.com}}

\author{Carlos Schat\\
        CONICET and Departamento de Fisica, FCEyN, Universidad de Buenos Aires, Ciudad Universitaria,
        Pab. 1, (1428) Buenos Aires, Argentina\\
Department of Physics and Astronomy, 
Ohio University, Athens, Ohio 45701, USA\\
        E-mail: \email{carlos.schat@gmail.com}}

\abstract{
We discuss the application of the permutation group $S_N$ to a few problems in hadron physics.
In Ref.~\cite{Pirjol:2007ed} a method was proposed for matching a quark model Hamiltonian onto the 
effective Hamiltonian of the $1/N_c$ expansion, which makes use of the transformation properties
of the states and operators under $S_N$. This method is used in \cite{Pirjol:2008gd}
to obtain information about
the spin-flavor structure of the quark interaction Hamiltonian from the spectrum of the negative parity
$L=1$ excited baryons. Assuming the most general 2-body quark Hamiltonian, we derive two correlations
among the masses and mixing angles of these states which should hold in any quark model.
These correlations constrain the mixing angles, and can be
used to test for the presence of 3-body quark interactions. We find that the pure gluon-exchange model
is disfavored by data, independently of any assumptions about the hadronic wave functions.}

\FullConference{International Workshop on Effective Field Theories: from the pion to the upsilon \\
		February 2-6 2009\\
		Valencia, Spain}

\begin{document}

\section{Introduction}
\label{intro}
Quark models provide a simple and  intuitive picture of the physics of 
ground state baryons and their excitations \cite{De Rujula:1975ge,Isgur:1978xj}. 
An alternative description is provided by the $1/N_c$ expansion,
which is a systematic and model-independent approach to the study of 
baryon properties \cite{Dashen:1993jt}.
This program can be realized in terms of a quark operator expansion, 
which gives rise to a physical picture similar to the one of the 
phenomenological quark models, but is closer connected to QCD.  In this 
context quark models gain additional significance. 

The $1/N_c$ expansion has been applied both to the ground state and excited nucleons
\cite{Goity:1996hk,Pirjol:1997bp,Carlson:1998vx,Schat:2001xr}.
In the system of negative parity $L=1$ excited baryons this approach has 
yielded a number of interesting insights. At leading order in $O(1/N_c)$, 
the following predictions follow from the contracted $SU(4)_c$ symmetry:

\begin{itemize}

\item  The three towers \cite{Pirjol:1997bp,Pirjol:2003ye,Cohen:2003tb} predicted 
by ${\cal K}$-symmetry 
for the $L=1$ negative parity $N^*$ baryons, labeled by ${\cal K}=0,1,2$ with ${\cal K}$
related to the isospin $I$ and spin $J$ of the $N^*$'s by $I+J \ge {\cal K}\ge|I-J|$. 

\item The vanishing of the strong decay width  $\Gamma(N^*_{\frac12} \to [N \pi]_S)$
for $N^*_{\frac12}$ in the ${\cal K}=0$ tower, which provides a 
natural explanation for the relative suppresion of pion decays for the  
$N^*(1535)$  \cite{Pirjol:1997bp,Pirjol:2003ye,Cohen:2003tb}.

\item The order ${\cal O}(N_c^0)$ mass splitting of the $SU(3)$ singlets   
$\Lambda(1405)$ - $\Lambda(1520)$
in the $[\mathbf{70},1^-]$ multiplet \cite{Schat:2001xr}. 

\end{itemize}

The $1/N_c$ expansion for the excited nucleons has been extended also 
to the first subleading order in $1/N_c$ 
\cite{Goity:1996hk,Pirjol:1997bp,Carlson:1998vx,Schat:2001xr,Pirjol:2003ye,Matagne:2004pm}.

In a recent paper \cite{Pirjol:2007ed} we showed how to match an 
arbitrary quark model Hamiltonian onto the operators of the $1/N_c$ expansion,
thus making the connection between these two physical pictures.
This method makes use of the transformation of the states and operators
under $S_N^{\rm sp-fl}$, the permutation group of $N$ objects acting
on the spin-flavor degrees of the quarks. This is similar to the 
method discussed in Ref.~\cite{Collins:1998ny} for $N_c=3$ in terms of
$S_3^{\rm orb}$, the permutation group of 3 objects acting on the
orbital degrees of freedom.

The main result of \cite{Pirjol:2007ed} can be summarized as follows:
consider a two-body quark Hamiltonian $V_{qq} = \sum_{i<j} O_{ij} R_{ij}$,
where $O_{ij}$ acts on the spin-flavor quark degrees of freedom, and $R_{ij}$
acts on the orbital degrees of freedom. Then the hadronic matrix elements of
the quark Hamiltonian on a baryon state $|B\rangle$ contains only the projections
$O_\alpha$ of $O_{ij}$ onto irreducible representations of $S_N$, the
permutation group of $N$ objects acting on the spin-flavor degrees
of freedom $\langle B |V_{qq}|B\rangle = \sum_\alpha C_\alpha 
\langle O_\alpha\rangle$. The coefficients $C_\alpha$ are related to
reduced matrix elements of the orbital operators $R_{ij}$, and are
given by overlap integrals of the quark model wave functions.

The explicit calculation in Ref.~\cite{Pirjol:2007ed} confirms the $N_c$ power 
counting rules of Ref.~\cite{Goity:1996hk,Carlson:1998vx}, in particular the 
leading order $O(N_c^0)$ contribution to the mass coming from the 
spin-orbit interaction $\vec s\cdot \vec l$, and confirms in a direct way the prediction 
of the breaking of the $SU(4)$ spin-flavor symmetry at leading order in $N_c$
\cite{Goity:1996hk}.
The calculation in Ref.~\cite{Pirjol:2007ed} confirms that the 
nonrelativistic quark model with gluon mediated quark interactions displays
the same breaking phenomenon. 

Another important conclusion following from the $S_N$ analysis is that 
operators depending on excited and core quarks are indeed required for a correct 
implementation of the $1/N_c$ expansion, in contrast to the approach of
Ref.~\cite{Matagne:2006dj} which does not include such operators. 

Any particular model of quark interactions, e.g the one-gluon 
exchange model (OGE) 
\cite{De Rujula:1975ge}, or the Goldstone boson exchange model (GBE)
\cite{Glozman:1995fu}, predicts a distinct hierarchy among the
coefficients $C_\alpha$  of the $1/N_c$ expansion. This prediction can be 
used to discriminate among models by confronting it against the observed
values of the coefficients.
 
In a recent paper \cite{Pirjol:2008gd} we used the $S_N$ approach to study 
the predictions of the quark model with the most general 2-body quark interactions, and to
obtain information about the spin-flavor structure of the quark interactions
from the observed spectrum of the $L=1$ negative parity baryons. This talk
summarizes the main results of this paper.

\section{The most general two-body quark Hamiltonian}
\label{Sec:Hamiltonian}

The most general 2-body quark interaction Hamiltonian in the constituent
quark model can be written in generic form as $V_{qq} = \sum_{i<j} V_{qq}(ij)$
with
\begin{eqnarray}
\label{2}
V_{qq}(ij) &=& \sum_k f_{0,k}(r_{ij}) O_{S,k}(ij) + 
f_{1,k}^a(r_{ij}) O_{V,k}^a(ij)  + 
f_{2,k}^{ab}(r_{ij}) O_{T,k}^{ab}(ij)\,, 
\end{eqnarray}
where $O_{S}, O_V^a, O_T^{ab}$ act on spin-flavor, and $f_k(r_{ij})$ are 
functions of $r_{ij} = |{\bf r}_i - {\bf r}_j|$. Their detailed form is unimportant
for our considerations. $a,b=1,2,3$ denote spatial indices.

\begin{table*}
\caption{The most general two-body spin-flavor quark interactions and their projections 
onto irreducible representations of $S_3$, the permutation group of three objects acting
on the spin-flavor degrees of freedom. $C_2(F)=\frac{F^2-1}{2F}$ is
the quadratic Casimir of the fundamental representation of $SU(F)$. }
\label{table_general}
\begin{tabular}{|c|c||c|c|}
\hline
Operator & ${\cal O}_{ij}$ & $O_S$ & $O_{MS}$   \\ 
\hline
Scalar & 1 &  1  & $-$  \\
              & $t_i^a t_j^a$ & $T^2 - 3C_2(F)$ & $T^2 - 3 t_1 T_c - 3C_2(F)$ \\
%              & $O_1, O_1-3O_2$ \\
              & $\vec s_i \cdot \vec s_j$ & $\vec S^2 - \frac94$ & $\vec S^2 - 3\vec s_1\cdot \vec S_c - \frac94$ \\
%              & $O_2+2O_3, O_2-O_3$ \\
              & $\vec s_i \cdot \vec s_j t_i^a t_j^a$ &  $G^2 - \frac94 C_2(F)$ & $3g_1G_c - G^2 + \frac94 C_2(F)$ \\
%              & $\frac{F}{4}O_1+\frac12 O_2+O_3$, \\
% & & & & $ O_1-(3+\frac{4}{F}) O_2+\frac{4}{F}O_3$ \\
\hline
Vector (symm) & $\vec s_i + \vec s_j$  & $\vec L\cdot \vec S$ & $3\vec L\cdot \vec s_1 - \vec L \cdot \vec S$ \\
%              & $O_4+O_5, 2O_5-O_4$  \\
   & $(\vec s_i + \vec s_j) t_i^a t_j^a$ & 
               $\frac12 L^i \{G^{ia}, T^a\} - C_2(F) L^i S^i$ 
               & $2 \frac{1-F}{F} L^i S_c^i + L^i g_1^{ia} T_c^a + L^i t_1^a G_c^{ia}$ \\
%               & $O_6+O_7+\frac{F-1}{2F}O_4, O_6+O_7-2\frac{F-1}{F} O_4$ \\
Vector (anti) & $\vec s_i - \vec s_j$  & $-$  & $3\vec L\cdot \vec s_1 - \vec L \cdot \vec S$ \\
%              & $2O_5-O_4$ \\
              & $(\vec s_i - \vec s_j) t_i^a t_j^a$ & $-$ & $L^i g_1^{ia} T_c^a - L^i t_1^a G_c^{ia}$ \\
%              & $O_6 - O_7$ \\
\hline
Tensor (symm) & $\{s_i^a , s_j^b\}$ & $L_2^{ij}\{S^i, S^j\}$ & $3L_2^{ij} \{ s_1^i, S_c^j\} - L_2^{ij} \{S^i, S^j\}$ \\
%              & $O_8 + 4 O_9, O_8-2O_9$ \\
   & $\{s_i^a , s_j^b\} t_i^c t_j^c$ & 
                $L_2^{ij} \{ G^{ia}, G^{ja}\}$ & 
                $L_2^{ij} g_1^{ia} G_c^{ja} - \frac{F-1}{4F}  L_2^{ij} \{ S_c^i, S_c^j\}$ \\
%              & $ \frac{F-1}{2 F} O_8 + 4O_{10}, \frac{F-1}{F}  O_8-4O_{10}$ \\
Tensor (anti)  & $[s_i^a , s_j^b]$ & $-$  & 0 \\
                 & $[s_i^a , s_j^b] t_i^c t_j^c$      & $-$  & 0 \\
\hline
\end{tabular}
\end{table*}

We list in Table~\ref{table_general} a complete set of spin-flavor 2-body 
operators with all possible Lorentz structures allowed by the orbital angular 
momentum $L=1$.
Columns 3 and 4 of Table~\ref{table_general} list the projections of 
the spin-flavor operators $O_{S}, O_V^a, O_T^{ab}$ onto the irreducible
representations of the $S_3$ permutation group, computed as explained in 
Ref.~\cite{Pirjol:2007ed}. 
The representation content depends on the symmetry of $O_{ij}$ under the 
permutation $[ij]$: the symmetric operators $O_{ij}$ are decomposed as 
$S+MS$, and antisymmetric $O_{ij}$ as $MS+A$. 

The symmetric $S$ projection depends only on quantities acting on the entire hadron 
$S^i, T^a, G^{ia}$, while the mixed-symmetric $MS$ operators depend on operators
acting on the core and excited quarks. 
We express them in a form commonly used in the application of the $1/N_c$ expansion 
\cite{Carlson:1998vx}, according to which their matrix elements are understood to be 
evaluated on the spin-flavor state $|\Phi(SI)\rangle$ constructed as a tensor product of an 
excited quark with a symmetric core with spin-flavor $S_c=I_c$.
The antisymmetric operators contain also an $A$ projection; its 
orbital matrix element vanishes for $N_c=3$ because of T-invariance 
\cite{Pirjol:2007ed,Collins:1998ny}, such that these operators do not contribute, 
and are not shown in Table~\ref{table_general}.

The orbital matrix elements yield factors of 
$L^i, L_2^{ij} = \frac12 \{L^i, L^j\} - \frac13\delta^{ij} L(L+1)$, which are 
the only possible structures which can carry the spatial index. 

From Table~\ref{table_general} one finds that the most general form of the
mass operator in the presence of 2-body quark interactions is a linear 
combination of 10 operators
\begin{eqnarray}\label{10Ops}
& & O_1 = T^2\,,\,\, O_2 = \vec S_c^2\,,\,\, O_3 = \vec s_1\cdot \vec S_c\,,
\,\, O_4 = \vec L\cdot \vec S_c\,,\,\, O_5 = \vec L\cdot \vec s_1\,,\,\,
O_6 = L^i t_1^a G_c^{ia}\,,\\
& & O_7 = L^i g_1^{ia} T_c^a\,,\,\,
O_8 = L_2^{ij} \{ S_c^i, S_c^j\} \,,\,\,
O_9 = L_2^{ij} s_1^i S_c^j \,,\,\, O_{10} = L_2^{ij} g_1^{ia} G_c^{ja} \,.
\nonumber
\end{eqnarray}
This gives the most general form of the hadronic mass operator 
of the negative parity $L=1$ states allowing only 2-body quark operators.

\section{Correlations}

The $L=1$ quark model states include the following SU(3) multiplets: 
two spin-1/2 octets $8_\frac12, 8'_\frac12$, two spin-3/2 octets $8_\frac32, 8'_\frac32$,
one spin-5/2 octet $8'_\frac52$, two decuplets $10_\frac12, 10_\frac32$ and two singlets
$1_\frac12, 1_\frac32$. States with the same quantum numbers mix, and we define
the relevant mixing angles in the nonstrange sector as
\begin{eqnarray}
\left\{
\begin{array}{cc}
N(1535) & = \cos\theta_{N1} N_{1/2} + \sin\theta_{N1} N'_{1/2}\\
N(1650) & = -\sin\theta_{N1} N_{1/2} + \cos\theta_{N1} N'_{1/2}\\
\end{array}
\right.\,,\qquad
\left\{
\begin{array}{cc}
N(1520) & = \cos\theta_{N3} N_{3/2} + \sin\theta_{N3} N'_{3/2}\\
N(1700) & = -\sin\theta_{N3} N_{3/2} + \cos\theta_{N3} N'_{3/2}\\
\end{array}
\right.
\end{eqnarray}
%and analogous for the $J=3/2$ states with the replacements 
%$(N(1535),N(1650),N_{1/2},N'_{1/2},
%\theta_{N1}) \to (N(1520),N(1700),N_{3/2},N'_{3/2},\theta_{N3})$.

It turns out that the 11 coefficients $C_{0-10}$ contribute to the mass
operator of the negative parity $N^*$ states only in 9 independent
combinations: $C_0,C_1 - C_3/2, C_2+C_3,
C_4, C_5, C_6, C_7, C_8+C_{10}/4, C_9 - 2C_{10}/3$.
This implies the existence of two universal 
relations among the masses of the 9 multiplets plus the two mixing angles, which must
hold in any quark model containing only 2-body quark interactions.

\begin{figure}[t!]
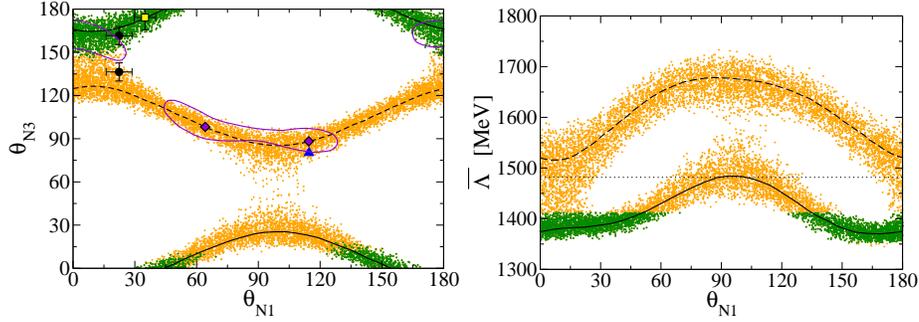

\begin{center}
\includegraphics[width=6.0cm]{fig1a.eps}
\includegraphics[width=6.0cm]{fig1b.eps}
\end{center}
\caption{Left: correlation in the $(\theta_{N1}, \theta_{N3})$ plane in the
quark model with the most general 2-body quark interactions. 
Right: prediction for the spin-weighted $\bar \Lambda$ mass in the 
SU(3) limit as a function of the $\theta_{N1}$ mixing angle, corresponding
to the two solutions for $\theta_{N3}$. The green points correspond to
$\bar \Lambda = \bar \Lambda_{\rm exp} - (100\pm 30)$ MeV, with
$\bar \Lambda_{\rm exp}=1481.7\pm 1.5$  MeV.}
\label{fig:OBE}
\end{figure}

The first universal relation involves only the nonstrange hadrons, and requires only 
isospin symmetry. It can be expressed as a correlation among the two mixing
angles $\theta_{N1}$ and $\theta_{N3}$ (see Fig.~\ref{fig:OBE} left)
\begin{eqnarray}\label{OBErelation}
&& \frac{1}{2} (N(1535) + N(1650)) + \frac{1}{2}(N(1535)-N(1650))
(3 \cos 2\theta_{N1} + \sin 2\theta_{N1}) \\
&& - \frac{7}{5} (N(1520) + N(1700)) + (N(1520) - N(1700))
\Big[ - \frac{3}{5} \cos 2\theta_{N3} + \sqrt{\frac52} \sin 2\theta_{N3}\Big] \nonumber \\
&& = -2 \Delta_{1/2} + 2 \Delta_{3/2} - \frac{9}{5} N_{5/2} \nonumber\,. 
\end{eqnarray}
This correlation holds also model independently in the $1/N_c$ expansion,
up to corrections of order $1/N_c^2$, since for non-strange states the 
mass operator to order $O(1/N_c)$ \cite{Carlson:1998vx,Schat:2001xr} is generated by
 the operators in Eq.~(\ref{10Ops}).
An example of an operator which violates this correlation is $L^i g^{ja} \{ S_c^j\,,
G_c^{ia}\}$, which can be introduced by 3-body quark forces. 

On the same plot we show also the values of the mixing angles obtained in several 
analyses of the $N^*\to N\pi$ strong decays and $N^*$ hadron masses.
The two black dots correspond to the mixing angles 
$(\theta_{N1}, \theta_{N3})=(22.3^\circ,136.4^\circ)$ and
$(22.3^\circ,161.6^\circ)$ obtained from a study of the strong decays in 
Ref.~\cite{Goity:2004ss}. The second point is favored by a $1/N_c$ analysis of
photoproduction amplitudes Ref.~\cite{Scoccola:2007sn}.
%DP should we add also the IK point $(31.7^\circ,173.6^\circ)$ ?
The yellow square corresponds to the
values used in Ref.~\cite{Carlson:1998vx, Schat:2001xr} 
$(\theta_{N1}, \theta_{N3})=(35.0^\circ,174.2^\circ)$, 
and the triangle gives the angles corresponding to the solution $1'$ in the large $N_c$ 
analysis of Ref.~\cite{Pirjol:2003ye}
$(\theta_{N1}, \theta_{N3})=(114.6^\circ,80.2^\circ)$. 
All these determinations (except the triangle) are compatible with the
ranges $\theta_{N1} = 0^\circ-35^\circ, \theta_{N3} = 135^\circ-180^\circ$. 
They are also in good agreement with the  
correlation Eq.~(\ref{OBErelation}), and provide no evidence for the presence of 3-body quark 
interactions. 

The second universal relation expresses the spin-weighted SU(3) singlet mass
$\bar \Lambda = \frac16  (2\Lambda_{1/2} + 4 \Lambda_{3/2})$ 
in terms of the nonstrange hadronic parameters
\begin{eqnarray}\label{LamAve}
\bar\Lambda &=& \frac16(N(1535) + N(1650)) + \frac{17}{15} (N(1520)+N(1700))
- \frac35 N_{5/2}(1675) - \Delta_{1/2}(1620) \\
& & - \frac16 (N(1535) - N(1650)) (\cos 2 \theta_{N1} + \sin 2\theta_{N1} )
 +
(N(1520) - N(1700)) 
(\frac{13}{15} \cos 2 \theta_{N3} - \frac13 \sqrt{\frac52} \sin 2\theta_{N3})\,.\nonumber
\end{eqnarray}
The rhs of Eq.~(\ref{LamAve}) is plotted as a function of $\theta_{N1}$ in 
the right panel of Fig.~\ref{fig:OBE}, where it can be compared against the experimental
value $\bar\Lambda = 1481.7\pm 1.5$ MeV. 
Allowing for SU(3) breaking effects $~\sim 100$ MeV, 
this constraint is also compatible with the range for $\theta_{N1}$ obtained above 
from direct determinations of the mixing angles.

Combining the Eqs.~(\ref{OBErelation}) and (\ref{LamAve}) gives a 
determination of the mixing angles from hadron masses alone, in contrast to
their usual determination from $N^*\to N\pi$ decays.
The green area in 
 Fig.~\ref{fig:OBE} shows the allowed region for $(\theta_{N1},
\theta_{N3})$ compatible with a positive SU(3) breaking correction in 
$\bar\Lambda$ of $100\pm 30$ MeV. One notes a good agreement between this 
determination of the mixing angles and that from $N^*\to N\pi$ strong decays.

\section{Spin-flavor structure of the quark interactions}

We derive next constraints on the spin-flavor structure of the quark interaction,
which can discriminate between models of effective quark interactions. There are two
popular models used in the literature. 
The first model is the one-gluon exchange model (OGE) \cite{De Rujula:1975ge}
which includes operators in Table~\ref{table_general} without isospin dependence. 
Expressed in terms of the  operator basis $O_{1-10}$ this gives the constraints
\begin{eqnarray}\label{OGEconst}
OGE: \qquad C_{1} = C_{6} = C_{7} = C_{10} = 0\,.
\end{eqnarray}

An alternative to the OGE model is the Goldstone boson exchange model (GBE) 
\cite{Glozman:1995fu}. 
In this model quark
forces are mediated by Goldstone boson exchange, and the quark Hamiltonian
contains all the operators in Table~\ref{table_general} which contain the flavor
dependent factor $t_i^a t_j^a$. The coefficients of the hadronic Hamiltonian $C_i$
satisfy the constraints ($F=3$ is the number of light quark flavors)
\begin{eqnarray}\label{OBEconst}
GBE: \qquad C_{1} = \frac{F}{4} C_{3}\,,\quad C_{5} = C_{9} = 0\,.
\end{eqnarray}

We would like to determine the coefficients $C_i$,
and compare their values with the predictions of the two models 
Eqs.~(\ref{OGEconst}), (\ref{OBEconst}). 
As mentioned, only 9 combinations of the 11 coefficients can be
determined from the available data: $C_0,C_1 - C_3/2, C_2+C_3,
C_4, C_5, C_6, C_7, C_8+C_{10}/4, C_9 - 2C_{10}/3$.
In particular, as the 
coefficients of the spin-orbit interaction terms $C_{4-7}$ can be determined,
we propose to use their values to discriminate between different models of quark interaction.

The values of $C_{4-7}$ can be compared with the hierarchy expected in each model.
In the OGE model the flavor-dependent operators have zero coefficients 
$C_{6,7} \sim 0 \ll |C_{4,5}|$,  while in the GBE model the spin-orbit interaction of the 
excited quark vanishes $C_5\sim 0 \ll |C_{4,6}|$. 

\begin{figure}[t!]
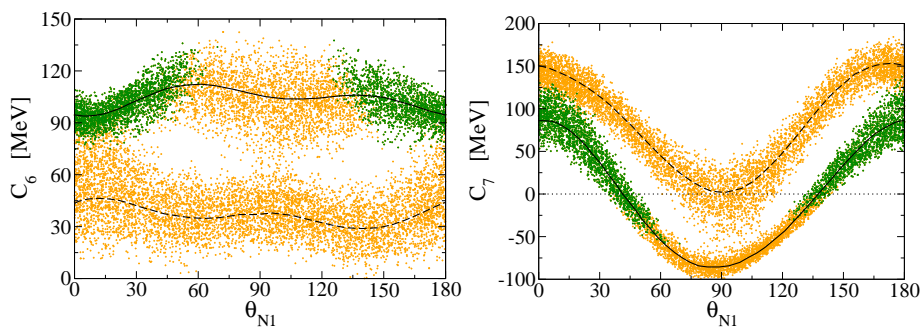

\begin{center}
\includegraphics[width=6.0cm]{fig2a.eps}
\includegraphics[width=6.0cm]{fig2b.eps}
\end{center}
\caption{The coefficients of the spin-orbit operators $C_{6,7}$ as functions
of the mixing angle $\theta_{N1}$, in the quark model with the most general
2-body interactions. 
The green area is obtained by
imposing the $\bar\Lambda$ constraint.}
% Eq.~(\ref{LamAve}). }
\label{fig:c4567}
\end{figure}

The coefficient $C_5= 75.7\pm 2.7$ MeV is fixed by the 
$\Lambda_{3/2}-\Lambda_{1/2}$ splitting \cite{Schat:2001xr}. This indicates the presence of
the operators $s_i \pm s_j$ in the quark Hamiltonian, which 
is compatible with the OGE model.
A suppression of the coefficients $C_{6,7}$ would be further evidence for the OGE model.
We show in Fig.~\ref{fig:c4567} the coefficients of the spin-orbit operators 
$C_{6,7}$ as functions of $\theta_{N1}$. 
Within errors small values for $C_7$ are still allowed, however no suppression
is observed for $C_6$. This indicates the presence of the operators
$(s_i \pm s_j) t^a_i t^a_j$ in the quark Hamiltonian. 
These results show that the quark Hamiltonian is a mix of the OGE and GBE
interactions. 

In the pure OGE model Eq.~(\ref{OGEconst}) the 7 nonvanishing coefficients $C_i$
can be determined from the 7 nonstrange $N^*,\Delta^*$ masses 
(assuming only isospin symmetry but no specific form of the wave functions).
This fixes the mixing angles, and the $\Lambda_{3/2}-\Lambda_{1/2}$ splitting,
up to a 2-fold ambiguity. The allowed region for mixing angles is shown as the 
violet region in Fig.~\ref{fig:OBE} left, and the central values as diamonds
$(\theta_{N1},\theta_{N3})=(64.2^\circ, 98.2^\circ), (114.5^\circ, 88.2^\circ)$.
Note that they are different from the angles obtained in 
the Isgur-Karl model $(31.7^\circ, 173.6^\circ)$ in Refs.~\cite{Isgur:1978xj,Chizma:2002qi,TBP}.

The violet region near $\theta_{N1} \sim 0$ is consistent with the
determinations from strong decays and from the SU(3) universal relation
Eq.~(\ref{LamAve}), but is ruled out by the prediction for the $\Lambda$ splitting,
in agreement with the non-zero value of $C_6$ that can be read off from
Fig.~\ref{fig:c4567}. This implies that the pure OGE model is disfavored
%DP
\footnote{Note that this argument neglects possible long-distance contributions to the 
$\Lambda$ splitting, due to the proximity of the $\Lambda(1405)$
to the $KN$ threshold. Such threshold effects are not described by the quark
Hamiltonian Eq.~(\ref{2}), and their presence could invalidate the prediction of
the $\Lambda$ splitting in the OGE model.}. 

\section{Conclusions}

We discussed a few applications of the permutation group $S_N$ to the study of baryonic
properties in the quark model. The applications are based on a simple result: the spin-flavor
contents of the mass operator is directly related to the projections of the spin-flavor
part of the quark interaction onto irreducible representations of $S_N$. Using this result,
any quark Hamiltonian can be matched onto the effective Hamiltonian of the $1/N_c$
expansion.

Following Ref.~\cite{Pirjol:2008gd}, we discussed the predictions of the most general 2-body quark Hamiltonian for
the spin-flavor structure of the negative parity $L=1$ excited baryons, without 
making any assumptions about the orbital hadronic wave functions. We derive two
universal correlations among masses and mixing angles, which constrain the mixing angles,
and can test for the presence of 3-body quark interactions. In addition, we derive
constraints on the spin-flavor structure of the quark forces from the observed
spectrum, and conclude that the gluon-exchange model is disfavored by data, 
independently on any assumptions about the hadronic wave functions.

\end{document}